# Quantum-Matter Heterostructures

**H. Boschker and J. Mannhart[1]**

h.boschker@fkf.mpg.de

office-mannhart@fkf.mpg.de

[1] corresponding author

Max Planck Institute for Solid State Research

Heisenbergstrasse 1

70569 Stuttgart, Germany

**Abstract**

Combining the power and possibilities of heterostructure engineering with the collective and emergent properties of quantum materials, quantum-matter heterostructures open a new arena of solid-state physics. Here we provide a review of interfaces and heterostructures made of quantum matter. Unique electronic states can be engineered in these structures, giving rise to unforeseeable opportunities for scientific discovery and potential applications. We discuss the present status of this nascent field of quantum-matter heterostructures, its limitations, perspectives, and challenges.

**Table of contents**

Keywords: Electronic correlations, Epitaxy, Quantum materials, Emergence, Defects

## 1.    Introduction

"*Even though incredibly rapid development has led to the present IC, a planar structure consisting of literally millions of circuits known as the chip, only a handful of materials is used, and of these mainly silicon.*" (1).

A momentous development has started in solid-state science that we expect will provide the basis for breakthroughs in science and applications of condensed matter for decades: Complex heterostructures such as superlattices, that previously were fabricated from standard semiconductors, simple metals or insulators, are now being grown from an increasing number of elements of the periodic table and compounds thereof. Figure 1 provides arbitrarily selected examples. The quality of some of these heterostructures is becoming comparable to that of semiconductor superlattices. Many heterostructures are grown with atomic precision and, in some cases, with abrupt interfaces. Well-defined, unit-cell-thick quantum wells are frequently achieved.

Heterostructures of quantum matter are particularly interesting systems. In this review we understand quantum matter to be defined by quantum effects that generate phenomena which surpass incoherent or mean-field behavior, and often are collective and emergent. Canonical quantum materials are the heavy-fermion compounds, the high-$T_c$ cuprates, the iridates, correlated organic compounds, and the iron pnictides. Unprecedented effects can occur if quantum materials are stacked, packed in quantum wells, brought into contact at interfaces, or altered by control parameters such as strain and electric fields imposed by gates. The phenomena thus induced are unforeseeable in their breadth and complexity. The development of quantum-matter heterostructures is exemplified by the development of transition-metal oxide heterostructures and the intriguing behavior of transition-metal oxide interfaces. These are the quantum-matter heterostructures to which presumably the greatest research efforts are being devoted. We discuss such systems, but also consider other material classes that are currently being explored, such as van-der-Waals heterostructures and electrostatically assembled heterostructures. To keep the contribution focused, we concentrate on heterostructures such as complex multilayers, superlattices, and their respective interfaces, rather than on single *p-n* junctions, Schottky contacts, and related structures.

We attempt to provide an overview of this young scientific field and shed light on the enormous possibilities it may generate. Several excellent reviews on subfields of quantum-matter heterostructures have already been written, including overviews and reviews of oxide heterostructures and oxide interfaces (see, *e.g.*, (2-29)), organic heterostructures (30), van-der-Waals heterostructures (31; 32), and electrostatically assembled heterostructures (32). We do not intend to update or duplicate these insightful reviews, but instead attempt to identify key issues of a scientific field that is extremely broad, characterized by complex questions, and is being explored by a wide spectrum of scientific activities. Our contribution can therefore not be complete; it omits some topics to better illuminate issues of a general character. The limited format of this article forces us to neglect several highly relevant contributions to this exciting field of heterostructures and interfaces involving quantum matter, for which we apologize in advance.

## 2.    Heterostructures grown from conventional materials

For many decades, heterostructures and superlattices have been used with outstanding success in science and applications. Innumerable dielectric multilayers serve, for example, as Bragg stacks in dielectric mirrors (33). Semiconducting heterostructures and superlattices, see, *e.g.,* (1; 34-40) are being applied in electronics and quantum optics (41). Many of them generate two-dimensional electron gases for the exploration of the quantum-Hall effects (42-44), host quantum wells in semiconductor lasers (45), or embed two-dimensional (2D) drain-source channels used, for example, in high-electron-mobility transistors of cell phones (46). Metallic heterostructures (see, *e.g.*, (47; 48)) are utilized, for example, as giant magnetoresistance devices in spintronics (12; 49-51) and in x-ray and neutron mirrors (52). These multilayers have been spectacularly successful because they enable functionalities that could not be obtained otherwise. What is behind this success? One key feature is that these structures can be grown to excellent levels of quality. The fact that the majority of the compounds applied are rather simple materials facilitates the growth. To grow such structures, high-quality substrates (*e.g.*, Si or GaAs wafers) are available. In addition, all these heterostructures rely on rather simple crystal structures and electronic properties. In cases such as optical coatings, the superlattice periods are thick on an atomic scale (typically tens of nanometers and larger), which makes them easier to fabricate. GaAs in particular is a good-natured compound. Its band gap can be engineered by alloying with Al, In, Ga, P, and other elements while keeping the lattice constants in the window required for the high-quality epitaxial growth of multilayers. Equally important, GaAs can be grown stoichiometrically by molecular-beam epitaxy (MBE) and metalorganic chemical vapor deposition using an overpressure of the highly volatile group V component, As (38). In that case, the defect chemistry of GaAs layers is given by thermodynamics and system purity. Nevertheless, in other technologically relevant semiconductor systems nature is not always accommodating. Nitride epitaxy, for example, is challenging because of the lack of high-quality substrates. Modern multijunction solar cells provide another example for multilayers that are challenging to grow, because these structures comprise materials for which growth problems are often difficult to master (see, *e.g.*, (53-55)).

In the heterostructures described above, the electronic behavior of the materials is well understood. Indeed, due to their single-particle, mean-field, rigid-band behavior, the electronic properties of these heterostructures can be engineered using band diagrams. Successful examples of such engineering are remote-doping to enhance the mobility of 2DEGs and the design of sub-bands in quantum cascade lasers. After an initial rocky phase in the 1950s (56), the single-particle approach became highly successful and valuable as a basis of heterostructure engineering. This prompted H. Kroemer to note in 2001 that, "If, in discussing a semiconductor problem, you cannot draw an Energy-Band-Diagram, this shows that you don't know what you are talking about, <...>. Nowhere is this more true than in the discussion of heterostructures, and much of the understanding of the latter is based on one's ability to draw their band diagrams—and knowing what they mean." (57)

But there is still more behind the spectacular properties of these multilayers. Consider, for example, that neutron-mirror superlattices are grown over large areas with Angstrom thickness precision (52), or that 2DEGs in GaAs have mobilities exceeding $3 \times 10^7$ cm$^2$/Vs with electronic mean-free paths of hundreds of micrometers (58; 59). How could this be achieved? In these cases the physics of growth and material properties is rather well understood. A few issues still remain open, referring, for example, to accurate values of band offsets or to the nature of scattering defects that determine the current record values for the mobilities. Although not enabling a routine fabrication of high-quality heterostructures, the science of semiconductor heterostructure fabrication has led to a good understanding of heterostructure growth. In particular in the early years of MBE, science benefitted from a productive cross-fertilizing interaction with commercial semiconductor technology. This good understanding of the growth of thin films with relatively simple crystal structures allowed heterostructures to be optimized in experimental studies. Nevertheless, this view is presumably simplistic and deceptive, as problems tend to seem obvious in hindsight. In this context it is worth noting that, from the first proposal of semiconductor heterostructures, it took half a century to achieve the current state of the art, with substantial progress taking place in the 1970s and 80s (1; 34; 60; 61).

### 3. A vision of quantum-matter heterostructures and their perspectives

*… imagine, then, a vast landscape (62)*

Regarding the freedom of design, quality, and achievements of heterostructures from standard dielectrics, semiconductors, and simple metals, one may conclude that their successes are restricted to materials of their kind. It is surprising then that, despite this common-sense understanding, new families of heterostructures grown from broad material classes hitherto unused for heterostructures are now being developed with remarkable success.

These families include heterostructures based on transition-metal oxides (2-29), van-der-Waals heterostructures obtained by stacking 2D materials (31; 32), electrostatically assembled heterostructures (32; 63; 64), and organic superlattices (30). Indeed, we discern a trend toward growing multilayers using elements from throughout the periodic table, opening up an ever-increasing choice of material combinations and stacking sequences. Scientifically exciting, many heterostructures are composed of quantum materials or produce them. It goes without saying that, for practical reasons, some elements and almost all of the possible compounds will obviously never be utilized, yet the scientific community is expanding the material space of heterostructures so steadily that we reflect in this article on the ultimate limit of heterostructures grown by benefitting from the complete material palette offered by the periodic table. This expansion of the material space is a grandiose, singular undertaking, for there is only one such table. What possibilities will arise and what difficulties might we encounter as we use more and more of the available elements to create quantum materials, often with the desire for atomic precision? In this contribution we address the state of the art, limits, and prospects of this endeavor.

There are good reasons why now is the time that the material space used for heterostructures is rapidly expanding. This development was boosted by the science-motivated interest in fabricating heterostructures from complex oxides, in particular from transition-metal oxides (65; 66). The epitaxial growth of complex oxides was spurred in the late 1980s and 1990s by the discovery of high-$T_c$ superconductors and the subsequent growth of cuprate films, superlattices (see, *e.g.,* (3)), Josephson junctions (see, *e.g.,* (67; 68)), and field-effect devices (see, *e.g.,* (69; 70)). It was accelerated by concomitant advances in oxide MBE, sputtering, and pulsed-laser deposition (PLD). Since then, four key developments have occurred: (i) a breakthrough in our understanding of the need for substrates terminated with single atomic planes and the development of preparation processes to achieve them (71; 72), (ii) the ability to monitor film growth by reflection high-energy electron diffraction (RHEED) even in relatively high partial pressures of oxygen (73-77), (iii) the use of ozone to grow oxide films by MBE (see, *e.g.,* (78-80)), and (iv) improvements made in the stoichiometry control of the deposited films (see, *e.g.,* (2; 81)), including the development of oxide hybrid MBE (82; 83). We suspect that the widespread growth of such heterostructures provided a tailwind to the efforts to grow also other, non-oxide-based complex multilayers such as the transition-metal dichalcogenides. Intriguingly, heterostructures can also be assembled manually from exfoliated sheets of two-dimensional materials (31). Also this technique is based on advances in the growth of high-$T_c$ superconductor films, in which adhesive tapes were used to cleave $Bi_2Sr_2CaCu_2O_{8+x}$ crystals to prepare tunnel junctions (see, *e.g.,* (84)). The exfoliation of graphene from graphite and the experience in handling these sheets opened the doors to the assembly of heterostructures incorporating other two-dimensional materials such as BN, $WSe_2$, $MoS_2$, and phosphorene.

Imagine for a moment that heterostructures could be grown by atomic-layer design by choosing basic sets from the vast portfolios of available structures and electronic properties available. Could we reproduce only the established solid-state physics and materials science in these samples? What is the exciting potential of these multilayers? Heterostructures are already allowing spectacular and qualitatively new electronic properties to be generated in mean-field semiconductor systems (85). The expansion of this material space to include quantum matter widens the categories of heterostructure electron systems in many directions. For illustration, Box 1 compares selected key properties of heterostructures of standard semiconductors with those of transition-metal oxide heterostructures. The phenomena listed tend to be coupled and mixed. For example, the lattice structure may control correlations, which may determine doping behavior, which in turn possibly couples back to the lattice.

Quantum matter provides many opportunities to engineer unprecedented electronic states. Coupling various phases in heterostructures may lead to new electronic properties. This can be realized both *in-plane* across self-assembled or patterned heterostructures and in the growth direction of the multilayers. For example, such couplings have been explored to enhance the transition temperature of superconductors (see, *e.g.*, (29; 86-89)). This approach makes use of the possibility given by quantum-matter heterostructures to spatially separate the mobile charge carriers and the coupling interactions, such that both can be optimized independently (87), (29). Indeed, non-superconducting electron systems may also be optimized by spatially separating the electron systems and their interactions. In addition, functionalities may become coupled. Quantum-matter heterostructures are therefore being explored with success as a means to generate, for example, multiferroic behavior (90-94). Interfaces may not only couple different order parameters and thereby achieve new effects, but they may also provoke phase transitions. Phases may be induced that already exist in this materials class in other parts of the phase diagram (for example by changing the doping at interfaces), and also completely new phases may be engineered, because basic electronic parameters such as the correlation parameters may change at the interfaces (9; 95). Already heterostructures of nickelates have been grown and analyzed to artificially mimic the band structure of the high-$T_c$ cuprates (96-99). Thus, the growth of quantum-matter heterostructures provides new degrees of freedom and a toolset to tailor and create materials, phases, effects, and functionalities that nature would not make on her own.

We are just at the beginning of a long development to use the one periodic table we have to assemble heterostructures with atomic control. Merely by the vast possibilities and surprising discoveries that will be made, the exploration of quantum-matter heterostructures will continue to be a burgeoning, highly rewarding field of science for decades to come. By the very nature of creating new materials and electron systems, this science will lead to many applications, the most important of which cannot yet be foreseen. Applications that seem promising today and are currently being explored include (i) the use of collective effects such as phase transitions (23; 100), electron-electron coupling, and negative capacitances to fabricate new field-effect transistors (FETs) (101), (ii) the use of heterostructures to promote catalytic behavior (see, *e.g.*, (102)) with a special emphasis on energy storage and conversion (water splitting, $CO_2$ reduction) (103; 104), and (iii) the design of materials with optimized work functions for energy conversion. The fabrication of multilayers involving high-$T_c$ cuprates for the fabrication of superconducting cables on the basis of the coated-conductor technology (105) is particularly advanced. The successful deposition of multilayers in kilometer length with well-aligned grain structures offers proof that the heterostructures of quantum matter are also promising candidates for large-scale applications.

**Box 1:**

**Basic Properties of Quantum-Matter Heterostructures**

**- Crystal structures**

In quantum-matter heterostructures, crystal structures are **usually more complex** compared to heterostructures based on, say, simple metals or semiconductors. **Structural distortions and defects** may be more relevant and complicated (*e.g.*, octahedra distortions and bond-angle changes that alter electronic and magnetic coupling and yield unusual magnetic structures), and also susceptibilities may be much larger and nonlinear (*e.g.*, $\varepsilon_r$ of $SrTiO_3$). Altering the crystal structure, for example by applying pressure or by adjusting the interlayer rotation of van-der-Waals heterostructures, provides degrees of freedom with possible effective control of heterostructure properties.

**- Electron systems**

Besides **partially filled 3*d* orbitals**, **partially filled 4*f* shells**, which are characterized by localized moments, and the **5*d* orbitals** that lead to heavy-fermion systems become available. Heavy nuclei may cause relativistic electronic effects, especially spin-orbit coupling.

Electron systems may be **correlated**; many-body effects can give rise to complex phases, order parameters and functionalities; electrons form liquids rather than gases (106);

electronic structures (band structures, correlation parameters) may change with doping, and new collective electronic states of matter and emergent properties occur (see. *e.g.*, (9; 16; 107)). As electron systems are frequently more localized than they are in semiconductors, defects may act in a more localized manner. In transition-metal heterostructures, large gradients of the chemical and electronic potentials may exist, which may drive chemical and structural reconstructions.

**Local effects** (*e.g.*, Coulomb repulsion $U$, Jahn-Teller distortions, octahedra physics) **and itinerant electron behavior** (characterized, *e.g.*, by double exchange), are relevant. Indirect exchange processes are common; effective masses are frequently larger and more dispersive than in semiconductors; mobilities are correspondingly lower (see below). Dirac materials have zero effective mass at the Dirac points.

Microscopic phenomena (*e.g.,* Coulomb repulsion, crystal field, elastic strain, Hund's exchange, orbital bandwidth, spin-orbit coupling) frequently show rather **small differences in energy scales**. Charge, spin, orbital and lattice **degrees of freedom exist**, and often **several interactions may be competing**, yielding competing ground states and phase diagrams. **Phase transitions** are therefore ubiquitous and systems often react sensitively to external parameters.

**Doping** (*e.g.,* in perovskites possible on *A, B*, O sites) has complex effects beyond altering charge density. Doping may change bond angles, defect properties, orbital couplings, and may induce phase transitions. Electronic effects of field-effect doping differ from chemical doping. Carrier densities can be much higher ($10^{22}/cm^3$) than in conventional semiconductors, potentially allowing for smaller devices. The electron systems therefore need higher polarizations to switch, *e.g.*, in FETs.

**Two-dimensional electron systems** tend to be more confined, less mobile, and correlated. Quantum wells can be made as thin as one atomic layer. Two-dimensional electron systems may couple to lattice functionalities. Active layers may be close to sample surfaces, and thus be more readily accessible to experiments, such as photoemission (108) and tunneling (109). Note that low-temperature mobilities of $7 \times 10^4$ cm$^2$/Vs, $1.4 \times 10^5$ cm$^2$/Vs and $>10^6$ cm$^2$/Vs have been achieved in LaAlO$_3$–La$_{1-x}$Sr$_x$MnO$_3$–SrTiO$_3$ (110), γ-Al$_2$O$_3$–SrTiO$_3$ (111), and in ZnO/(Mg,Zn)O multilayers (112), respectively, $> 2 \times 10^5$ cm$^2$/Vs in free-standing graphene (113) and $10^7$ cm$^2$/Vs in graphene on surfaces of bulk graphite (114).

- **Interfaces** may change electronic states and correlation parameters. Interfaces may generate new electronic phases, couple different order parameters and orbital rearrangements (9; 95; 115; 116).

- **Functionalities**

A **large phase space** of lattice configurations and electron systems provides freedom to design new Fermi surfaces and electron systems with novel excitations and quasiparticles.

More **functionalities** are available, such as enhanced work function tuning, superconductivity, magnetism, sensor functions, triggering of phase transitions, nanoferroic effects, intrinsic amplification, and resistive switching. Ionic conduction and storage are more relevant. New mechanical, thermal, and thermoelectronic properties may be achieved.

Please note also the lists comparing typical parameters of semiconductor and oxide interfaces given in the supplementary of (9) and in (13).

## 4. Current developments in the field of quantum-matter heterostructures

We are just beginning to assemble heterostructures with atomic control from an increasingly broad spectrum of our periodic table. Epitaxially grown quantum-matter heterostructures are ubiquitous today, as illustrated by Fig. 1. The figure shows a series of transmission-electron-microscopy images of heterostructures taken from literature. The examples include heterostructures from a selection of complex oxides (64; 117-127), graphene and BN (128), Co:$BaFe_2As_2$ and $SrTiO_3$ (129), Sc, $B_4C$, and Cr (130), and $Sb_2Te_3$ and GeSbTe (131) in addition to heterostructures from GaAs and AlAs (132) and GaN and AlN (133) reproduced for comparison. These images also demonstrate the ubiquity of epitaxially grown quantum-matter heterostructures. Moreover, atomically flat interfaces have been realized in a wide range of materials and are being used to study a broad spectrum of physical and chemical phenomena. Excellent summaries of these phenomena are given in prior reviews (2-32). We note an ever increasing diversity of the heterostructures. The materials range expends, different crystal structures and crystal orientations are being used, and epitaxial strain is systematically controlled. Thicknesses can be controlled down to the single unit cell level (134) and this way atomically thin conducting sheets have been achieved (122). Furthermore, the heterostructures have been integrated into electronic circuits (135). Parallel to these achievements, theoretical proposals for exotic properties occurring at quantum-matter heterostructures consider an increasing variety of chemical elements, crystal orientations, stacking sequences, strain states, etc. (14; 136-138). In this way, the design of quantum heterostructures benefits from the total phase space of orbital configurations and of the possibilities for functional tuning provided by quantum matter.

FeSe–SrTiO$_3$ heterostructures provide a scientifically highly interesting example for quantum-matter heterostructures. In the simplest case this heterostructure consists of a single unit-cell-thick layer of FeSe grown by molecular beam epitaxy on SrTiO$_3$ (001). A two-dimensional electron system resides in the FeSe layer, and it superconducts at rather high temperatures. Both scanning tunneling spectroscopy and angle-resolved photoemission spectroscopy measurements show a large superconducting gap of up to 20 mV (4.2 K), which persists to ~65 K (139-141). A superconducting transition was also observed by transport measurements, and was found to depend on variations of the sample structure. A $T_c$ of 15 K has been reached by capping the FeSe with an FeTe layer (142), a $T_c$ of 40 K by using ion-gated electric double layer transistor devices (143), and a $T_c$ of 109 K is claimed on the basis of an *in situ* four-probe measurement of a bare film using mechanical contacts (144). Bulk FeSe, in contrast, has a $T_c$ below 10 K that can be increased up to approximately 30-50 K by doping or pressure (145; 146). The FeSe–SrTiO$_3$ system therefore provides a clear case in which heterostructuring enhances $T_c$ compared to bulk samples. We note that FeSe-SrTiO$_3$ heterostructures are being further explored to provide answers to further fundamental questions. It is, for example, intriguing to ask whether $T_c$ can be enhanced even more by creating FeSe–SrTiO$_3$ superlattices (147). Also the mechanisms that result in the increase of $T_c$ of FeSe–SrTiO$_3$ bilayers ask for clarification. Possible contributing factors are electron doping of the FeSe by oxygen vacancies in the SrTiO$_3$ (140), the epitaxial strain that enforces a tetragonal symmetry in the FeSe (148), coupling of the electrons in FeSe to phonons in the SrTiO$_3$ (149), and the large dielectric susceptibility of SrTiO$_3$ (150). These factors highlight the complex, interwoven character of the structural and electronic properties of quantum-matter heterostructures. The FeSe–SrTiO$_3$ system also demonstrates the possibilities the heterostructures offer for the creation of unique, novel electron systems.

At present, several trends and developments, overarching materials classes, and deposition methods, are shown by the development of the heterostructures in the expanding materials space. The complexity of the materials science and physics of 2D electron systems and interfaces tends to increase, and surprising electronic, magnetic, and optical behaviors are observed. The heterostructures are patterned into the nano-regime through electron beam lithography (135; 151-153) and scanning-probe-based writing (154). Self-organization and self-assembly yield excellent 3D structures (90; 94). Mesoscopic phenomena, quantum transport, and one-dimensional and zero-dimensional systems are being explored with growing interest (85; 122; 152; 155; 156). Furthermore, structures with non-standard symmetry properties may possibly provide novel electronic properties, and an increasing number of samples is being designed and grown to induce frustration, exotic superconductivity, and tantalizing topological phenomena (24; 138; 157-159). Ion transport within the heterostructures is recognized to provide an additional degree of freedom. It has also been found that simultaneous electron and ion conduction offers new device possibilities (160-164).

The freedom given by nature to design and fabricate heterostructures does have its limits, even to those who do not shy away from growing toxic, radioactive, or excessively costly samples. The requirements on stability or metastability of the structures limits the material space to compounds that do not react destructively with each other or decompose, also under the growth conditions used for subsequent layers and during patterning. The current state of the art in heterostructure fabrication has been achieved on one hand by a continuous series of technical improvements. New growth techniques are being pioneered, and others continue to be advanced. These advances include better vacuum systems, high-purity sources including gas supplies, better lasers and optics, and increasingly powerful growth-monitoring equipment. Cluster tools use combinations of different growth techniques to deposit the layers of a stack. On the other hand, a better understanding of epitaxial growth processes benefitted the state of the art and enabled the growth of higher-quality heterostructures. The challenge of growing two materials $A$ and $B$ repeatedly on top of each other with atomically abrupt interfaces amounts to being able to nucleate a monolayer of the right phase of material $A$ on top of material $B$ and vice versa, and to being able to grow both materials with a two-dimensional growth mode.

We now discuss the challenge of nucleating material $B$ on top of $A$. With increasing complexity of the compounds, the need to control the formation of desired phases requires appropriate and sometimes challenging control of the growth parameters. If material $B$ consists of more than one chemical element, the desired phase can only be nucleated if it is at least meta-stable with respect to alternative phases of different stoichiometries. The nucleation of material $B$ is much easier when both materials have similar crystal structures, as is the case in almost all high-quality heterostructures. Indeed, the epitaxial stabilization provided during growth is a powerful tool to control the nucleation of the desired phases, even enabling stabilization of compounds that do not exist in nature. When growing complicated materials involving multiple elements, it is key to control accurately the stoichiometry of the arriving species in the deposition process. Unwanted phases are frequently triggered by small off-stoichiometries. Once the unwanted phases have been nucleated, they can generally no longer be converted.

The nature of the bonding between the two materials needs to be considered, too. The bonds range from weak van-der-Waals bonds between the sheets of two-dimensionally bonded materials to strong covalent and ionic bonds in three-dimensionally bonded materials. Van-der-Waals epitaxy is distinct from conventional epitaxy (165). Two-dimensional materials tend to grow layered parallel to the substrate surface. Owing to their weak bonding with the underlying layers, grains frequently nucleate with a spread of *in-plane* orientations. Therefore *in-plane* alignment is a challenge for van-der-Waals epitaxy. Control of the *in-plane* orientation, however, is possible in the epitaxy of two-dimensionally bonded materials by using appropriate surface reconstructions (166).

Next, we discuss two-dimensional growth. The standard theory of growth in thermodynamic equilibrium, based on the values of the surface and interfaces energies, that leads to Volmer–Weber, Frank–van der Merve, and Stranski–Krastanov growth does not provide clear guidance for the growth of complex heterostructures with atomically flat interfaces. Consider the case in which the surface energy of material $A$ is higher than that of material $B$ and that the interface energy is small. Then $B$ can be grown atomically flat on top of $A$, but $A$ will grow three-dimensionally on $B$. However, atomically flat interfaces are frequently needed for both sequences. There are ways to accomplish this. Surface modifications using either reconstructions or surfactants can be used to promote two-dimensional growth. Growth kinetic effects can also be exploited to achieve flat surfaces. These effects include destabilization of small islands, interval deposition, and controlling the energy of the arriving species.

Now we turn to the theoretical modeling of the heterostructures. Understanding the properties of quantum-matter heterostructures requires understanding the properties of the interfaces. This is by no means a trivial task. Typical parameters such as $U$, $J$, or $m^*$ can change their values at interfaces (9; 95). Moreover, the structural and electronic degrees of freedom tend to be strongly coupled, requiring many structural variations to be calculated. The complexity and large system size of the heterostructure imply the need for combining different models, including first-principles and phenomenological theory. Electronic correlations may be treated by first-principles methods such as dynamic mean-field theory (167-169), the functional renormalization group (170) and full-configuration-interaction quantum Monte Carlo (171). For describing emergent properties such as magnetism and superconductivity, phenomenological models are often applied.

Density functional theory (DFT) (172-174) is used most widely for calculating the properties of heterostructures (see *e.g.*, (175)). One of its main advantages is that it allows one to calculate both the atomic positions and the electronic structure. The comparison between theory and experiment is therefore possible at a variety of scales. These include the atomic and electronic structures and low-energy properties such as electric transport properties and magnetism. It is imperative that the atomic structure of the samples be understood together with low-energy properties. In this context, the development of better characterization tools for analyzing the heterostructures has yielded much progress (see, *e.g.*, (108; 176-189)). Improvements in scanning transmission electron microscopy and x-ray scattering techniques have provided unprecedented structural information. Several research groups are performing *in-situ* angle-resolved photoemission or tunneling measurements to reveal the intrinsic electronic structure. Knowing the atomic and electronic structures allows a direct comparison of experimental results and findings from first-principles theories. Model calculations can then be used to construct models describing such phenomena as electronic transport properties. Predicting the functional properties is the most challenging aspect of understanding heterostructures, and some of these challenges are described in the next chapter.

## 5.    Challenges and future research directions

We now look toward the future to consider key questions and challenges. Which fertile directions for research and applications of heterostructures can be identified today that will leverage the vast spectrum of compounds available? What are the dominant showstoppers preventing novel ideas for electron systems in heterostructures from being transformed into real samples that show the predicted properties? Although much effort has already been undertaken to this effect, many open research topics have yet to be addressed.

A key issue that prevents intriguing ideas from becoming reality is the occurrence of defects. The range of defects that occur in quantum-matter heterostructures is generally enlarged due to the increased chemical complexity of these systems. More defects are present, and more types of defects are possible. Great progress has been made in the past twenty years regarding the growth of heterostructures from complex compounds. Nonetheless there are open issues concerning their growth and the characterization and control of defects. Several issues, like surface reconstructions or controlling interface roughness, also relate to heterostructures from metals, dielectrics, and semiconductors.

Indeed, even with the progress already made, the fabrication of quantum-matter heterostructures will benefit at all stages of growth from a series of advances we consider likely to happen. Substrate quality and the availability of substrates with a large selection of lattice symmetries and constants have improved in recent years (190). Nevertheless, a choice of affordable, high-quality, large-area substrates or semiconductor wafers covered with a buffer layer system would boost the field. The same is true if termination procedures were available for more substrate materials and surface orientations. The control of interface quality and defects in heterostructures requires *in-situ* deposition and thus, for some material combinations, the use of cluster tools. Better reproducibility and defect control will be achieved by avoiding *ex-situ* sample fabrication steps such as the *ex-situ* termination of interfaces or the manual assembly of van-der-Waals bonded heterostructures. The accurate control of growth parameters, such as substrate surface temperature during growth as affected, *e.g.*, by infrared emission properties changing during film deposition, and the precise online control of film thickness are to be improved. The accuracy of controlling film thickness by RHEED is constricted by phase-shift effects, the precise determination of oscillation maxima during film growth and, for PLD, by the sizable number of adatoms arriving in the individual deposition pulses. Sample surfaces may affect active layers even if these are remote from the surface as evidenced by the effects seen at buried 2D electron liquids in $LaAlO_3$–$SrTiO_3$ interfaces (191-193). Surfaces therefore must be controlled. This is typically achieved by growing inert capping layers, which is done routinely when completing semiconductor heterostructures. Related effects may be generated during patterning. Patterning damage, produced for example by ion etching, may introduce trap states, conducting shunts, dead layers, and second phases.

With the increasing complexity of quantum-matter heterostructures, it may become ever more difficult for experimental studies to identify intrinsic phenomena amidst defect-induced effects. At present, limitations to sample quality are already given by defects generated by the substrates (dislocations, mosaic structure, surface steps, impurities, and non-stoichiometries) and by defects caused by the materials used for growth. Many rare earths and their compounds are available only with purities up to 99.99%. Another issue concerns, for example, defect–defect interactions. Defect clustering or compensation may change the properties of the defects qualitatively. We refer in this context to the work on $Sr_{n+1}Ti_nO_{3n+1}$ (129) layers in which the incorporation of $SrO_2$ shear planes succeeded in reducing the point-defect density in the material matrix, yielding record low microwave losses (see Fig. 1). Better analytical tools are needed for the characterization of heterostructures and their defects. For example, the currently available analytical tools are notoriously inadequate for measuring oxygen concentrations and point-defect densities in thin samples or at interface layers. Atomic positions and interdiffusion on the atomic scale are also difficult to quantify. Interpretation of scanning transmission electron microscopy (STEM) cross sections of interfaces, for example, requires the projection of possible growth steps to be known. Analytical tools are challenged by the small amount of material to be analyzed, in particular at interfaces and their defects. Progress made in aberration-corrected STEM and electron energy-loss spectroscopy is impressive, and there is great potential also in the imaging of light atoms. The complexity, however, of quantum-matter heterostructures and the sensitivity and emergent character of their electronic behavior is generating another class of problems: the analytical tools used to characterize these systems may alter their structure and behavior. The instruments may then provide information on electron systems with qualities that are different than the unperturbed ones. This is relevant for studies using electron irradiation (194-196), photoemission (197-201), and STEM (202).

In our opinion, the effects of defects and analytical tools on sample properties pose considerable challenges to the development of the field. Real-life heterostructures usually do not match the idealized Lego-like structures envisioned during design. Samples grown by different groups may differ in complex ways. Further differences may be induced if the samples are characterized by procedures that, as mentioned, change the sample properties. Adding to the complexity, defects may provoke a series of ramifications. Ionized point defects, for example, may induce changes in carrier density, screening properties, lattice distortions, and built-in potentials. This entanglement of electronic parameters is an intrinsic property of quantum matter and as such has to be accepted. This complexity, however, must be taken into account when analyzing data and comparing different samples, in particular if results of different measurement techniques applied to different samples are combined. Therefore it would be desirable if agreement could be reached as to what information should be provided in publications about sample properties and characterization. For example it would be helpful if data were commonly given about the composition, surface characterization performed by scanning force microscopy, x-ray diffraction, and whether the analysis altered the samples' properties.

The comparison between theory and experiment also poses challenges. First of all it is relatively easy for theorists to predict that new properties will occur at heterostructures that do not yet exist. For experimentalists, however, it takes substantial effort to grow such heterostructures. Typically, experimentalists will exert this effort only if the initial results are promising and will emphasize the results only if they were positive. This generates a feedback loop in which successful predictions are reported and predictions that are probably incorrect remain untested, leading to overly optimistic notions of our understanding of the heterostructures.

Second, the calculation of realistic defect densities is difficult. These calculations depend on whether defects are intrinsic or extrinsic to the ground state of the system. In the latter case, calculations performed for $T$=0 favor an absence of defects. However, defect-formation energies can be calculated and compared with defect-chemistry measurements at extreme conditions (high temperatures and low pressures). This can then lead to quantitative models of the defect densities at typical measurement conditions. Calculations of the effects of defects on the heterostructures' properties require large supercells.

An example of a heterostructure in which defects can be intrinsically stable in the ground-state is a system with a polar discontinuity. In that case, the artificially imposed boundary conditions at the interface result in an electrostatic energy cost. This energy can lead to the emergence of spectacular changes of the interface properties, but it will also favor the creation of defects that compensate the interfacial charge mismatch. For $LaAlO_3$–$SrTiO_3$ interfaces, for example, recent DFT calculations show that oxygen vacancies are more favorable at the $LaAlO_3$ surface (203-209), *i.e.*, the $LaAlO_3$–vacuum interface. Similarly, Sr vacancies can be expected at the $SrTiO_3$ side of the interface. It is possible that, for some interfaces, first-principles theories that allow for a free density of cation and anion vacancies at the interface and surface, respectively, will not predict a two-dimensional electron liquid at all, and will have the polar discontinuity mostly compensated by defects. In practice, however, heterostructures are not at chemical equilibrium. Defect densities can be influenced to some extent during growth and cooldown. Thus, it is sometimes even feasible to achieve defect densities smaller than the ones predicted from equilibrium conditions.

It is clear that, in a research arena as rich as quantum-matter heterostructures and in which the most important progress is yet to come, important research directions will evolve in the future as scientific progress leads to grand new challenges. But even today, research trends are discernible that go beyond the obvious, such as improving growth methods. Box 2 lists such research opportunities that we predict will be especially worthwhile to explore.

**Box 2:**

**Possible Directions of Future Research on Quantum-Matter Heterostructures**

**More and more elements** of the periodic table will be used. Film growers will have to deal with a further increasing number of elements and will find solutions for this.

Epitaxial growth of non-cubic crystal systems and epitaxial growth involving layers of different crystal symmetries will likely be used for **crafting new types of heterostructures and quantum matter**. Predictive atomistic growth and annealing models, for example, for MBE or PLD, would be highly helpful. Indeed, any improvements in growth simulation would foster experiments. The models must be able to consider growth temperatures, surface reconstructions, surface enegetics, non-equilibrium growth kinetics, the identity of diffusing surface species and their rates, and defects.

The **characterization and understanding of defects** will be a central area of research. The impact of defects on sample properties may be less detrimental in quantum-matter than in semiconductors because electron systems tend to be more localized. The impact could also be more virulent owing to emergent effects. The impact of the defects depends on which material system and properties are affected. A promising research topic is the use of emergent effects for defect mitigation.

Research will be devoted to allowing quantum-matter heterostructures to be grown on larger substrates such as Si wafers, and to increasingly integrate them with semiconductor technology. Interface-based structures may feature numerous 2D systems in parallel with adjustable coupling using both $p$ and $n$-doping. The increase of the mobilities of 2D systems will continue to be an exciting research field, whereas the record mobility values of semiconductors will be difficult or even impossible to achieve with many types of quantum matter. 3D-patterned heterostructures would open a new world of research and device applications of their own, but would have to be implemented without unacceptable sacrifices of layer quality.

We expect that **ionic transport** will gain relevance in future quantum-matter heterostructures. Ionic conduction and electron conduction in parallel as well as ionically active functionalities will offer new freedom for designing devices.

Quantum heterostructures offer numerous possibilities for generating **catalytic active surfaces**, benefitting from tuning the surface electron system by structural or electronic effects induced by layers underneath. Water splitting or $CO_2$ reduction are important research topics and applications of interest. **Work-function tuning** performed by similar principles may yield surface properties desired for applications, particularly in the field of green energy.

Correlation effects have potential for **device applications**. Examples include the use of collective electron effects to achieve transport of carriers of charge $n \times e$ with $n > 1$ in devices, and switching effects use intrinsic amplifications generated by switching of emergent electron properties or phase transitions.

**Coherent quantum-matter heterostructures** may be explored for use in research and in devices. Mini-bands are used with great success in semiconductor heterostructures; the degrees of freedom for mini-band engineering are even greater in quantum-matter heterostructures, for example to tailor the band structure of correlated materials. Lateral **structuring of quantum-matter heterostructures to quantum length scales** will open the door to achieving 1D and 0D systems based on quantum matter, such as artificial atoms.

For numerous quantum matter heterostructures accurate, four-terminal electrical measurements of transport perpendicular to the heterostructure planes are highly desirable. These are impeded, however, by sizable resistances of the compounds used to contact the active part of the device and by their contact resistances. Optimization of device configurations to measure with precision transport in c-direction would lead to new fields of research on quantum matter heterostructures.

**The exploration of nonequilibrium effects** of quantum-matter heterostructures, induced, *e.g.*, by pump-probe experiments or by quasiparticle injection, will generate a wealth of new phenomena.

The possibility to control lattice symmetries and stacking sequences in quantum-matter heterostructures, *e.g.*, by epitaxial growth, opens the road to create new materials characterized by **frustration**, for example in the spin or orbital degrees of freedom, and to fabricate electron systems characterized by **special topological properties**.

The wide variety of models developed for the understanding of quantum-matter heterostructures will increasingly be integrated into a **coherent theoretical framework** and thereby provide enhanced prediction accuracy.

## 6.  Summary and conclusions

Quantum-matter heterostructures are enticing for science and technology. These heterostructures involve a continuously increasing number of elements and compounds. Electron systems that can be achieved with them far extend the set of fundamental parameters of heterostructures based on silicon, III–V, II–VI, or IV–VI compounds, and other classic materials such that some of the fundamental limitations and restrictions of the semiconductor heterostructures do not apply to these heterostructures. Important progress has been made, yet a number of key issues must still be overcome to continue this progress.

Real-world heterostructures and films are far more complex than toy bricks. Owing to their complex microstructures and electronic properties, defects may provoke entangled effects. A meaningful comparison of the results obtained by a characterization technique *A* on a sample grown by group *B* with the results of a sample of group *C* analyzed with technique *D* may require a detailed understanding of the sample properties, the underlying physics, and the measurement procedures used. This issue is compounded by the fact that the field is broad and many different heterostructures can be grown from numerous materials, making it a challenge to stay abreast of the field. Numerous groups may follow different standards. It is therefore difficult to distill the intrinsic behavior of heterostructures from a variety of published datasets.

Although the quality of the quantum-matter heterostructures is catching up with non-quantum matter systems, the availability of high-quality, large-area substrates and purer source materials is desirable, as is an even better control of defects created during deposition. Developments of new growth processes are well underway. Fabrication of all heterostructures would benefit greatly from the availability of predictive growth models and analytical tools for defect characterization.

Successfully designing quantum-matter heterostructures will be feasible once the growth models are accurate on the required low-energy scales. The combination of theoretical methods, the consideration of larger supercells, and more detailed comparison to experiments will enhance the understanding of the heterostructures.

The great potential of heterostructures grown from many different compounds was already envisioned in the 1980s. Scientists endeavored to grow heterostructures such as artificially layered superconductors and exploratory high-$T_c$ superconducting devices. Since then, the state of the art has progressed much further than the proof-of-principle that such samples can be fabricated. The improvement in the fabrication of such heterostructures is thriving, but we still have a long way to go. The field is veritably exploding in width and depth. Science has just scratched the surface of an enormously fertile ground of great application potential and unimaginable limits.

**Acknowledgements**

The authors gratefully acknowledge helpful interactions with P. Böni, I. Bozovic, W. Braun, A. Brinkman, W. Dietsche, J. Eckstein, J. Falson, T. Kopp, B. Lotsch, K. Ploog, D.G. Schlom, and H. Takagi.

**Figure Caption**

*Figure 1*

Transmission electron microscopy images of heterostructures. A large variety of materials can be grown today into heterostructures, often with atomically abrupt interfaces. These heterostructures provide functionalities that are not available in bulk compounds. Adapted from the original publications (see references next to the images) with permission.

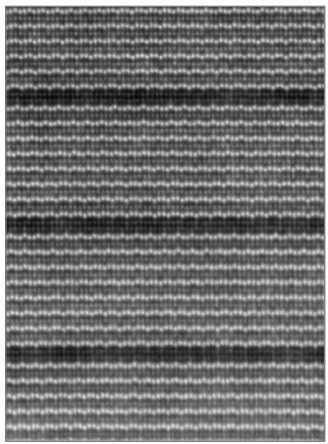

LuFeO$_3$ – LuFe$_2$O$_4$

2 nm

(117)

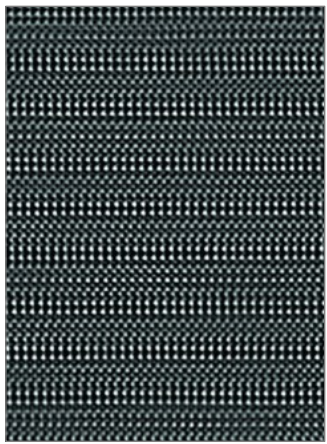

La$_{0.7}$Sr$_{0.3}$MnO$_3$ – SrRuO$_3$

3 nm

(119)

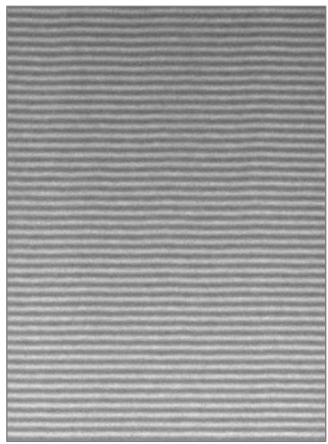

Sc – B$_4$C – Cr

20 nm

(130)

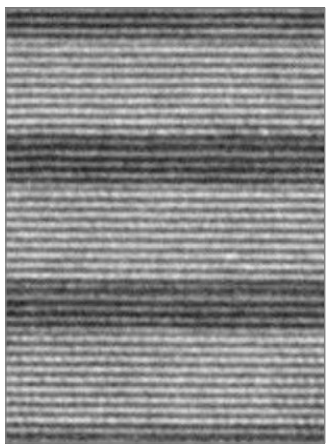

GaN – AlN

1 nm

(133)

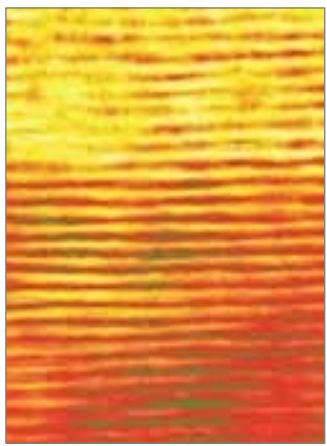

Graphene – BN

5 nm

(128)

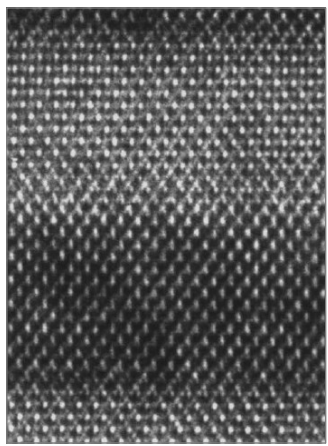

GaAs – AlAs

1 nm

(132)

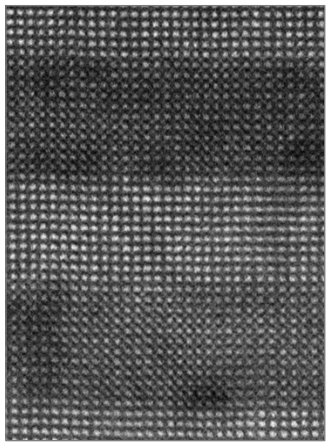

LaAlO$_3$ – SrTiO$_3$

3 nm

(118)

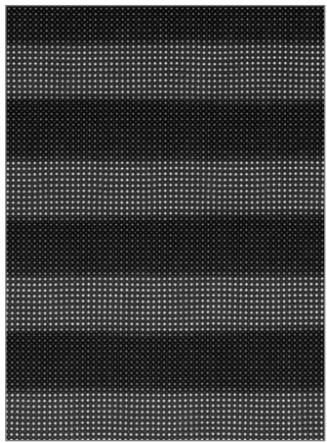

PbTiO$_3$ – SrTiO$_3$

4 nm

(120)

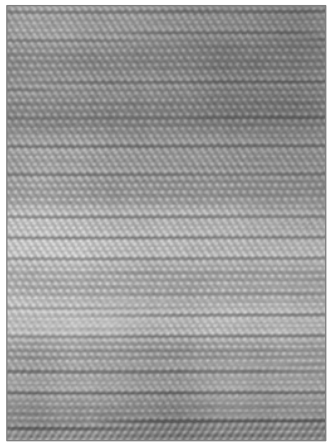

Sb$_2$Te$_3$ – GeSbTe

4 nm

(131)

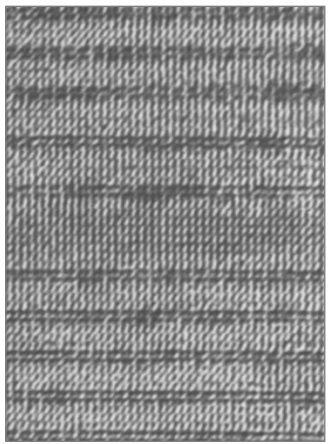

$Bi_2Sr_2Ca_7Cu_8O_{20} - Bi_2Sr_2CaCu_2O_8$

2 nm (121)

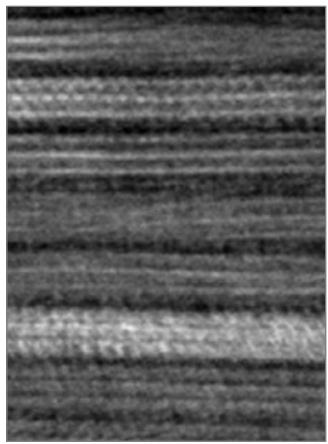

$Mn_2Al(OH)_6 - Ca_2Nb_3O_{10}$

2 nm (64)

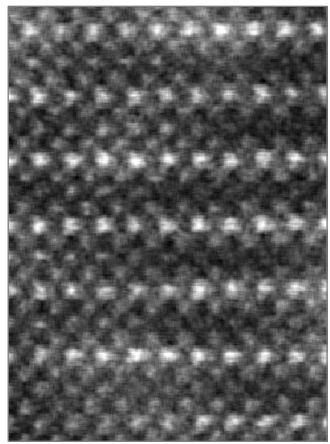

$SrIrO_3 - SrTiO_3$

1 nm (125)

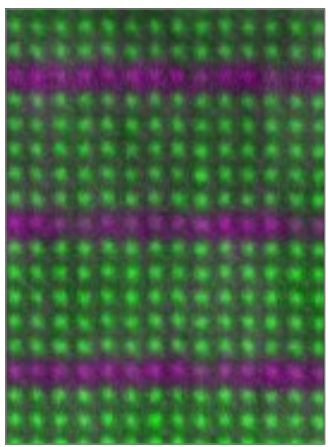

$SrRuO_3 - SrTiO_3$

1 nm (122)

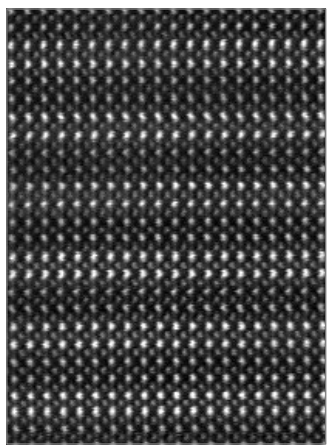

$LaMnO_3 - SrMnO_3$

2 nm (124)

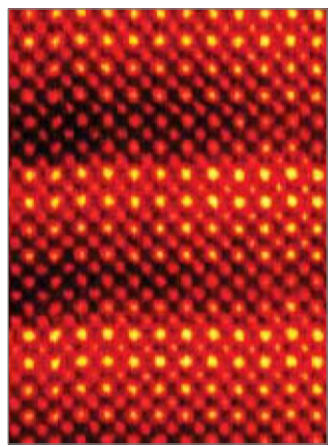

$BaTiO_3 - SrTiO_3 - CaTiO_3$

1 nm (126)

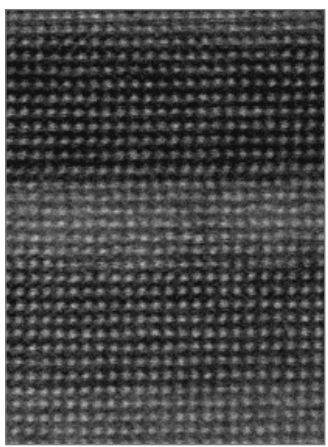

$LaMnO_3 - LaNiO_3$

2 nm (123)

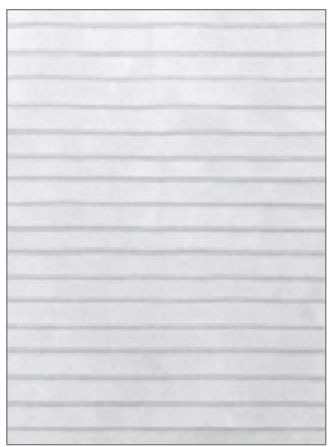

$Co:BaFe_2As_2 - SrTiO_3$

20 nm (129)

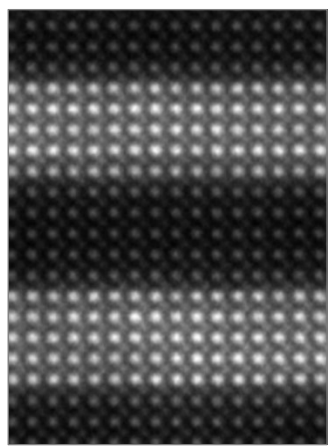

$La_{0.7}Sr_{0.3}MnO_3 - SrTiO_3$

1 nm (127)